\documentclass[twoside,twocolumn,9pt]{article}
\usepackage{extsizes}
\usepackage[super,sort&compress,comma]{natbib} 
\usepackage[version=3]{mhchem}
\usepackage[left=1.5cm, right=1.5cm, top=1.785cm, bottom=2.0cm]{geometry}
\usepackage{balance}
\usepackage{times,mathptmx}
\usepackage{sectsty}
\usepackage{graphicx} 
\usepackage{lastpage}
\usepackage[format=plain,justification=justified,singlelinecheck=false,font={stretch=1.125,small,sf},labelfont=bf,labelsep=space]{caption}
\usepackage{float}
\usepackage{fancyhdr}
\usepackage{fnpos}
\usepackage[english]{babel}
\addto{\captionsenglish}{%
  
}
\usepackage{array}
\usepackage{droidsans}
\usepackage{charter}
\usepackage[T1]{fontenc}
\usepackage[usenames,dvipsnames]{xcolor}
\usepackage{setspace}
\usepackage[compact]{titlesec}
\usepackage{hyperref}
\usepackage{siunitx}
\usepackage[capitalise]{cleveref}
\usepackage{subcaption}
\usepackage{amssymb}

\definecolor{cream}{RGB}{222,217,201}

\begin{document}

\pagestyle{fancy}
\thispagestyle{plain}
\fancypagestyle{plain}{

\fancyhead[C]{\includegraphics[width=18.5cm]{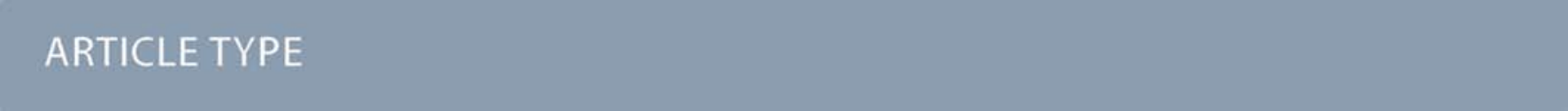}}
\fancyhead[L]{\hspace{0cm}\vspace{1.5cm}\includegraphics[height=30pt]{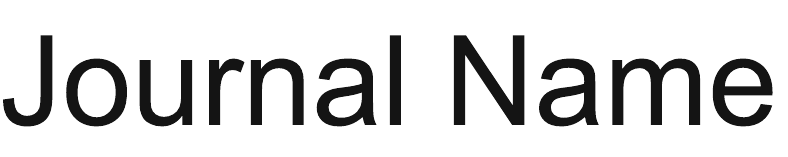}}
\fancyhead[R]{\hspace{0cm}\vspace{1.7cm}\includegraphics[height=55pt]{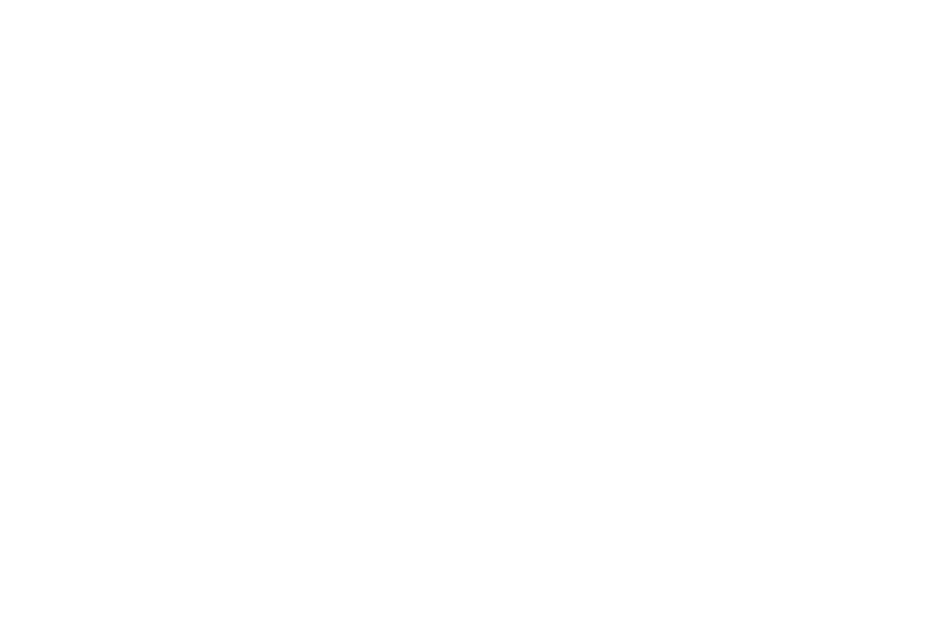}}
\renewcommand{\headrulewidth}{0pt}
}

\makeFNbottom
\makeatletter
\renewcommand\LARGE{\@setfontsize\LARGE{15pt}{17}}
\renewcommand\Large{\@setfontsize\Large{12pt}{14}}
\renewcommand\large{\@setfontsize\large{10pt}{12}}
\renewcommand\footnotesize{\@setfontsize\footnotesize{7pt}{10}}
\makeatother

\renewcommand{\thefootnote}{\fnsymbol{footnote}}
\renewcommand\footnoterule{\vspace*{1pt}%
\color{cream}\hrule width 3.5in height 0.4pt \color{black}\vspace*{5pt}} 
\setcounter{secnumdepth}{5}

\makeatletter 
\renewcommand\@biblabel[1]{#1}            
\renewcommand\@makefntext[1]%
{\noindent\makebox[0pt][r]{\@thefnmark\,}#1}
\makeatother 
\renewcommand{\figurename}{\small{Fig.}~}
\sectionfont{\sffamily\Large}
\subsectionfont{\normalsize}
\subsubsectionfont{\bf}
\setstretch{1.125} 
\setlength{\skip\footins}{0.8cm}
\setlength{\footnotesep}{0.25cm}
\setlength{\jot}{10pt}
\titlespacing*{\section}{0pt}{4pt}{4pt}
\titlespacing*{\subsection}{0pt}{15pt}{1pt}

\fancyfoot{}
\fancyfoot[LO,RE]{\vspace{-7.1pt}\includegraphics[height=9pt]{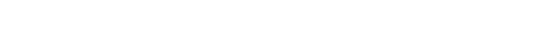}}
\fancyfoot[CO]{\vspace{-7.1pt}\hspace{13.2cm}\includegraphics{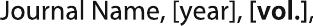}}
\fancyfoot[CE]{\vspace{-7.2pt}\hspace{-14.2cm}\includegraphics{RF}}
\fancyfoot[RO]{\footnotesize{\sffamily{1--\pageref{LastPage} ~\textbar  \hspace{2pt}\thepage}}}
\fancyfoot[LE]{\footnotesize{\sffamily{\thepage~\textbar\hspace{3.45cm} 1--\pageref{LastPage}}}}
\fancyhead{}
\renewcommand{\headrulewidth}{0pt} 
\renewcommand{\footrulewidth}{0pt}
\setlength{\arrayrulewidth}{1pt}
\setlength{\columnsep}{6.5mm}
\setlength\bibsep{1pt}

\makeatletter 
\newlength{\figrulesep} 
\setlength{\figrulesep}{0.5\textfloatsep} 

\newcommand{\topfigrule}{\vspace*{-1pt}%
\noindent{\color{cream}\rule[-\figrulesep]{\columnwidth}{1.5pt}} }

\newcommand{\botfigrule}{\vspace*{-2pt}%
\noindent{\color{cream}\rule[\figrulesep]{\columnwidth}{1.5pt}} }

\newcommand{\dblfigrule}{\vspace*{-1pt}%
\noindent{\color{cream}\rule[-\figrulesep]{\textwidth}{1.5pt}} }

\definecolor{ppp}{rgb}{0.71, 0.20, 0.54}
\definecolor{qqq}{rgb}{0.57, 0.17, 0.16}
\newcommand{\Max}[1]{
 {\color{ppp} #1 }
}
\newcommand{\Note}[1]{
 {\color{qqq} #1 }
}

\makeatother

\twocolumn[
  \begin{@twocolumnfalse}
\vspace{3cm}
\sffamily
\begin{tabular}{m{4.5cm} p{13.5cm} }

\includegraphics{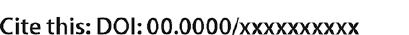} & \noindent\LARGE{\textbf{Capillary assemblies in a rotating magnetic field}} \\
\vspace{0.3cm} & \vspace{0.3cm} \\

 & \noindent\large{Galien Grosjean,$^{\ast}$\textit{$^{a}$} Maxime Hubert\textit{$^{a,c}$}, Ylona Collard\textit{$^{a}$}, Alexander Sukhov\textit{$^{b}$}, Jens Harting\textit{$^{b,e}$}, Ana-Sun\v{c}ana Smith\textit{$^{c,d}$} and Nicolas Vandewalle\textit{$^{a}$}} \\

\includegraphics{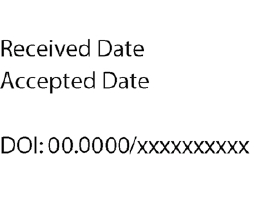} & \noindent\normalsize{Small objects floating on a fluid have a tendency to aggregate due to capillary forces.
This effect has been used, with the help of a magnetic induction field, to assemble submillimeter metallic spheres into a variety of structures, whose shape and size can be tuned.
Under time-varying fields, these assemblies can propel themselves due to a breaking of time reversal symmetry in their adopted shapes.
In this article, we study the influence of an in-plane rotation of the magnetic field on these structures.
Various rotational modes have been observed with different underlying mechanisms.
The magnetic properties of the particles cause them to rotate individually.
Dipole-dipole interactions in the assembly can cause the whole structure to align with the field. 
Finally, non-reciprocal deformations can power the rotation of the assembly.
Symmetry plays an important role in the dynamics, as well as the frequency and amplitude of the applied field.
Understanding the interplay of these effects is essential, both to explain previous observations and to develop new functions for these assemblies.} \\

\end{tabular}

 \end{@twocolumnfalse} \vspace{0.6cm}

  ]

\renewcommand*\rmdefault{bch}\normalfont\upshape
\rmfamily
\section*{}
\vspace{-1cm}


\footnotetext{\textit{$^{a}$~GRASP Lab, CESAM Research Unit, University of Li\`ege, B-4000 Li\`ege, Belgium; E-mail: ga.grosjean@uliege.be}}
\footnotetext{\textit{$^{b}$~Helmholtz Institute Erlangen-N{\"u}rnberg for Renewable Energy (IEK-11), Forschungs\-zentrum J{\"u}lich, F{\"u}rther Stra{\ss}e 248, 90429 N{\"u}rnberg, Germany.}}
\footnotetext{\textit{$^{c}$~PULS Group, Institute for Theoretical Physics and Cluster of Excellence: Engineering of Advanced Materials, Friedrich Alexander University Erlangen-N{\"u}rnberg, Cauerstra{\ss}e 3, 91058 Erlangen, Germany.}}
\footnotetext{\textit{$^{d}$~Group for Computational Life Sciences, Division of Physical Chemistry, Institute Ru\dj{}er Bo\v{s}kovic, Zagreb, Croatia.}}
\footnotetext{\textit{$^{e}$~Department of Applied Physics, Eindhoven University of Technology, P.O. box 513, NL-5600MB Eindhoven, The Netherlands.}}




\section{Introduction}

As a way of simplifying the handling of microscopic objects, self-assembly has been intensely studied in recent years~\cite{whitesides2002,whitesides2002b,zhang2009b,thorkelsson2015,boles2016,dou2017}.
In particular, supramolecular self-assembly has allowed to undertake an impressive number of complex tasks including low-Reynolds-number mixing~\cite{rida2004}, drug delivery~\cite{majedi2013}, assay of drugs~\cite{wu2008}, cargo transport~\cite{martinez2015}, microfabrication~\cite{chung2008}, controlled self-propulsion~\cite{grosjean2015}, photonics~\cite{kim2011} or sensoring~\cite{edel2013}.
By requiring no direct intervention, self-assembly can offer clever solutions at length scales, ranging from the nanometer to the centimeter, that are difficult to access using traditional fabrication techniques, motors and actuators.
One particular system uses magnetic spheres suspended at an interface to form reversible clusters called magnetocapillary self-assemblies~\cite{vandewalle2012,vandewalle2013}.
These assemblies differ from other magnetic floating crystals~\cite{golosovsky1999,wen2000,grzybowski2000,golosovsky2002} in that they can form outside of a confining potential.
In this case, the particles experience a pairwise lateral capillary attractive force~\cite{kralchevsky1994,vella2005} as well as a magnetic dipole-dipole repulsive force.
The resulting interaction potential resembles the Derjaguin-Landau-Verwey-Overbeek (DLVO) potential used in colloidal science and adsorption~\cite{ninham1991}, with a primary minimum that corresponds to the generally irreversible contact between the particles, and a secondary minimum that corresponds to an equilibrium distance where the particles can be reversibly trapped~\cite{vandewalle2012}.

\begin{figure}[t!]
\centering
\includegraphics[width=\linewidth]{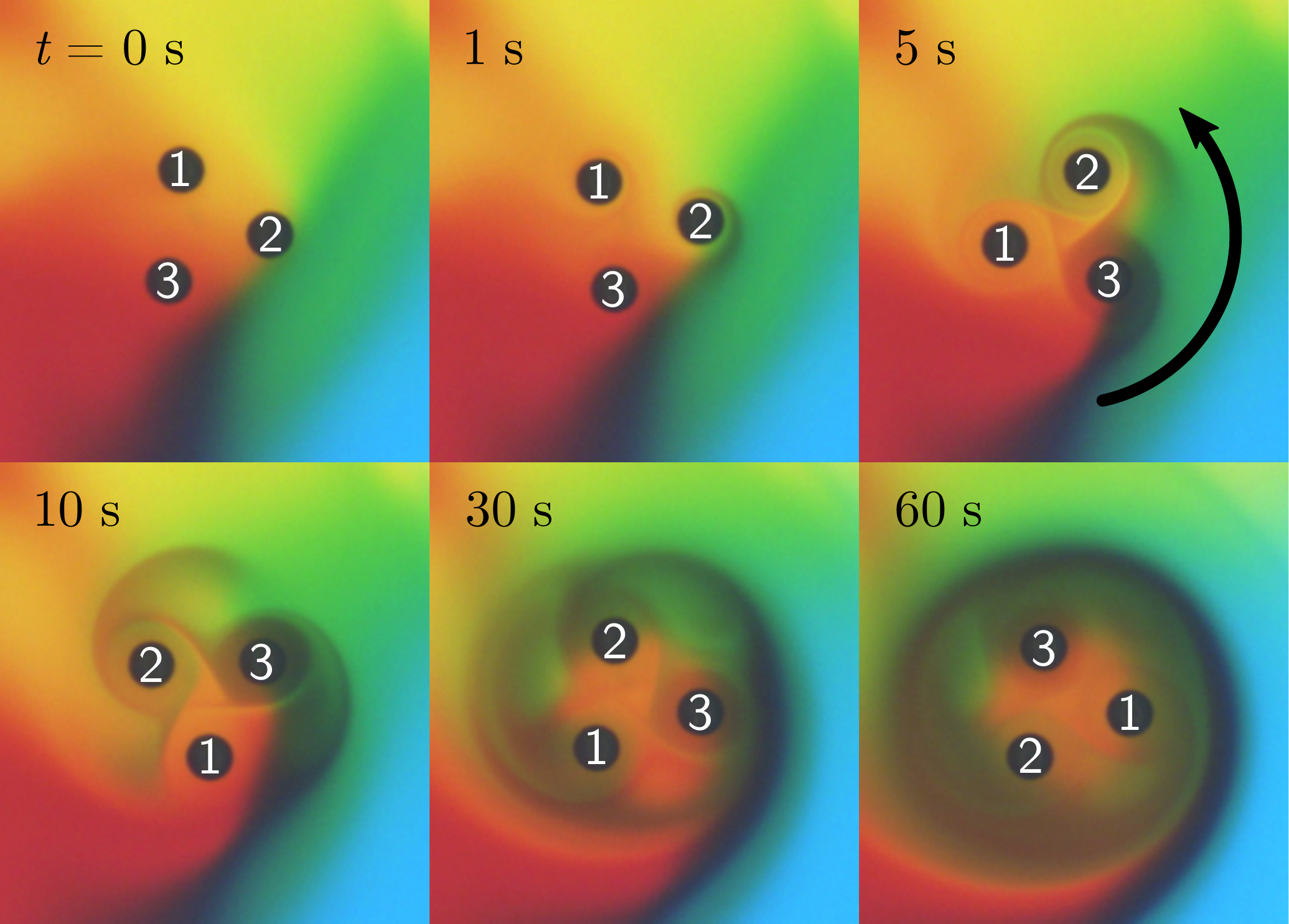}
\caption{
\textbf{Successive photographs of a triangular assembly.} --
A triangular assembly of \SI{500}{\micro\meter} particles is exposed to a constant vertical field and a horizontal field rotating in the plane of the interface at $f = \SI{1}{\hertz}$.
The assembly rotates with an angular speed $\Omega \approx \SI{0.3}{\radian\per\second}$.
Dyes allow us to see the fluid motion due to both the rotation of the whole assembly and the individual rotations of the particles.}
\label{ylo}
\end{figure}

The resulting structures can be made to perform several tasks, such as remote-controlled swimming~\cite{lumay2013,grosjean2015}, transport and fluid mixing~\cite{grosjean2018}.
They can also serve to study fundamental principles of low-Reynolds-number locomotion~\cite{grosjean2016,pande2017,sukhov2019}.
Compared to other interface-bound systems~\cite{grosjean2018b}, they do not use surface waves~\cite{snezhko2006,snezhko2011} or surface tension gradients~\cite{nakata2013,bassik2008,bormashenko2015,okawa2009} for locomotion.
Instead, the role of the interface is to provide the attractive force and confinement necessary to form the assembly.
The motion itself is powered by the non-reciprocal deformation sequence adopted by the assemblies when subjected to time-dependent magnetic fields.

However, several essential questions remain unanswered.
First, the mechanism behind the rotational motion of the individual particles is not well understood, despite being necessary to explain the non-reciprocal deformation required for propulsion~\cite{grosjean2015,chinomona2015}.
Secondly, the rotational dynamics of an assembly has never been experimentally studied, despite being used to change the swimming direction~\cite{grosjean2015} or locally stir fluid without turbulence~\cite{grosjean2018}.
For instance, \cref{ylo} shows the stirring of fluids by an assembly under a rotating field.
Dyes allow to visualize the flow, showing both a rotation of the individual particles and a slower rotation of the whole structure.
In general, a trio of rotating particles can cause complex dynamics in the Stokes regime, despite the fact that the Stokes equations are linear and time-reversible~\cite{lushi2015}.
This can be an advantage for micromixing, particularly if there are multiple length and time scales involved.
Finally, studying the influence of a rotating field on a magnetocapillary assembly can teach us more about the many swimming regimes observed~\cite{hubert2013} and the nontrivial influence of the oscillating field's amplitude, the field's frequency and the geometry of the assembly~\cite{grosjean2015,grosjean2018}.

\section{Experimental setup and protocol}

\begin{figure}[t]
\includegraphics[width=\linewidth]{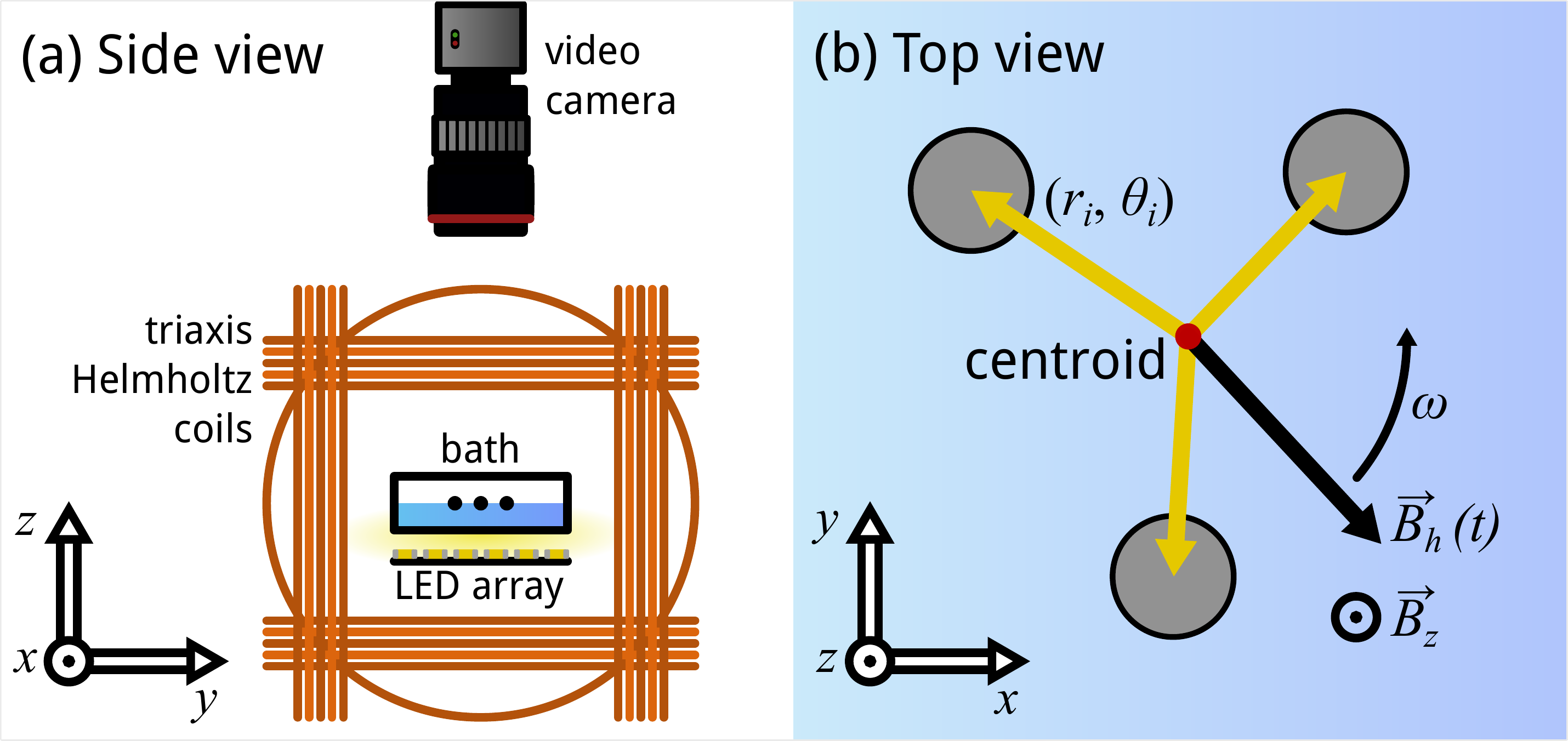}
\caption{\textbf{Experimental schematic.} --
(a) Side view of the experiment, showing the $x$, $y$ and $z$ coils, the water bath and the acquisition setup.
(b) Top view of an assembly of three particles defined by their coordinates ($r$,~$\theta$) in the referential of the center of mass.
They are exposed to a vertical magnetic field $B_z$ and a horizontal field $B_h$ rotating at an angular frequency $\omega = 2\pi f$.
}
\label{setup}
\refstepcounter{subfigure}
\label{setup:a}
\refstepcounter{subfigure}
\label{setup:b}
\end{figure}

A glass container filled with water is placed at the center of a tri-axis Helmholtz coil system, as illustrated in \cref{setup:a}.
A transverse Hall Effect probe was used to obtain the ratio between the amplitude of the magnetic field and the current applied in each coil, as well as measure the spatial uniformity of the field. An integrated triaxis magnetoresistive sensor was also used to test the orthogonality between the coils. As the coils are order of magnitudes larger than the assemblies, about \SI{50}{\centi\meter} in diameter, the nonuniformity of the magnetic field is negligible.
Currents in the $x$ and $y$ coils are injected using a computer-controlled multichannel arbitrary function generator and a pair of amplifier, while the $z$ coils are fed by a regulated DC power supply.

Metallic spheres are deposited on the water surface with a pair of titanium tweezers.
To minimize contamination, the tank is covered by a glass lid and the typical experimental runs are kept short, typically below an hour.
Before each experimental run, the tank and lid are cleaned, and the water and particles replaced. 
The bottom side of the lid is coated with a transparent conductive oxide and connected to earth to prevent the accumulation of electric charges.
A video camera fitted with a macro lens records the motion of the particles from the top.
An LED array illuminates the bath from below.
The particles are tracked using an algorithm based on circle Hough transforms.

The spheres used have a diameter $D$ of \SI{500}{\micro\meter}, are highly monodisperse, and are made of either steel or alloy steel, most commonly UNS G52986 and UNS S42000.
Although they are 7.8 times denser than water, they float thanks to surface tension.
Each particle is described by its polar coordinates $r_i$ and $\theta_i$ in the center-of-mass frame, as shown in \cref{setup:b}.
We will not consider displacements along the vertical direction, as they are typically very small.
Indeed, for the range of particle-to-particle distances used herein, we can estimate using the Laplace equation that vertical displacements are typically around \SI{1}{\percent} of $D$.

Under a magnetic induction field $\vec{B}$, the particles display a highly linear magnetization~\cite{lagubeau2016}.
This means that their magnetic moment is aligned with, and proportional to the field such that $\vec{m} \approx \chi V \vec{B} / \mu_0$ with $V$ the volume of the spheres and $\chi \approx 3$ their magnetic susceptibility.
We consider here that the particles are perfectly isotropic and display no residual magnetism, or remanence.
Indeed, the remanence measured in \cite{lagubeau2016} is about two orders of magnitude smaller than $\chi V B / \mu_0$ for typical values of $B$.
However, as can be seen in \cref{ylo}, the individual particles can rotate around their center.
This can only be explained by taking into account some anisotropy in the properties of the beads, as the induced component is always parallel to the external field.
Its origin will be discussed later on.
We can create the following ansatz for the magnetization of the particles
\begin{equation}
\vec{m} = \frac{\chi V \vec{B}}{\mu_0} + \vec{m}_0,
	\label{Eq:TranslBeads}
\end{equation}
where $\vec{m}_0$ is a vector of fixed magnitude and orientation with respect to the bead that accounts for the anisotropy in magnetic properties.
In general, though, the contribution of the remanence $\vec{m}_0$ is negligible.

The magnetic induction field used throughout this paper takes the form of a constant vertical field $\vec{B}_z$ and a horizontal field $\vec{B}_h$ of constant magnitude rotating parallel to the interface at an angular frequency $\omega = 2\pi f$.
The total magnetic field vector is therefore precessing around the $z$ axis at a frequency $f$.
We have
\begin{equation}
\begin{split}
\vec{B}_z &= B_z \vec{e}_z, \\
\vec{B}_h &= B_h \cos\left(\omega t\right) \vec{e}_x + B_h \sin\left(\omega t\right) \vec{e}_y,
\label{fields}
\end{split}
\end{equation}
and the amplitude of the vertical field is kept constant at a value $B_z = \SI{4.9}{\milli\tesla}$.
Changing $B_z$ would likely change the relative importance of the various effects discussed here, and therefore the frontiers of the regimes observed.
However, to limit the number of control parameters in the experiments, we chose to mainly vary the excitation frequency $f$ and amplitude $B_h$.

\section{Static assembly}

In the absence of a magnetic field, the particles aggregate thanks to a lateral capillary force, a phenomenon known as the Cheerios effect~\cite{kralchevsky1994,vella2015}.
Floating particles are surrounded by a meniscus, which in the case of a spherical particle of diameter $D$ is given by
\begin{equation}
\frac{z}{D} \propto K_0 \left(\frac{d}{l_c}\right),
\label{meniscus}
\end{equation}
where $l_c = \sqrt{\gamma / \rho g}$ is the capillary length that relates surface tension $\gamma$ and unit weight $\rho g$, and $K_0$ is the zeroth order modified Bessel function of the second kind.
This assumes small deformations compared to the capillary length.

In general, a solution of the Laplace equation cannot be found in the case of two particles.
However, if the particles are separated by a large distance $d$, one can assume that the profile is given by the sum of both individual menisci.
If a particle experiences a net vertical force, such as the resultant force of gravity and buoyancy, then an interaction potential arises from the product of such vertical force and the vertical displacement caused by the second particle.
In other words, the slope of the interface can cause the particle to rise or fall along the meniscus, leading to a horizontal displacement.
The potential energy associated with this interaction is given by
\begin{equation}
U_c = -2\pi\gamma q^2 K_0 \left(\frac{d}{l_c}\right),
\label{Uc}
\end{equation}
where $q = L \sin \Psi$ is called the capillary charge, with $L$ the radius of the contact line and $\Psi$ the meniscus slope angle.
This is assuming that the particles have the same capillary charge, which leads to an attractive force.
The capillary charge is a measurement of a typical deformation length.
We typically have $q \approx \SI{10}{\micro\meter}$ for a \SI{500}{\micro\meter} particle both from theory~\cite{vella2005,lagubeau2016} and from direct observations of the contact line.
The validity of the linear and superposition approximations has been experimentally tested in~\cite{lagubeau2016}.
These approximations reproduce the experimental findings quantitatively granted that the particles are separated by a distance $d \gtrsim 2D$, which is always the case in the present paper.

\begin{figure}[t]
\begin{subfigure}{0.49\linewidth}
	\includegraphics[width=\linewidth]{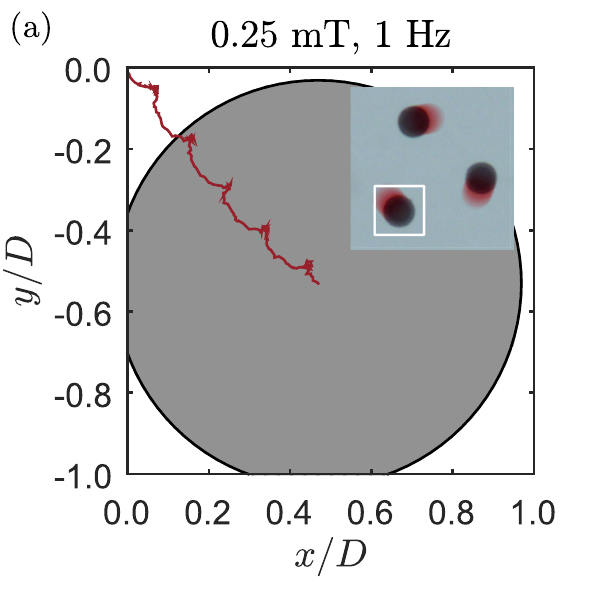}
	\refstepcounter{subfigure}
	\label{snaps:a}
\end{subfigure}
\begin{subfigure}{0.49\linewidth}
	\includegraphics[width=\linewidth]{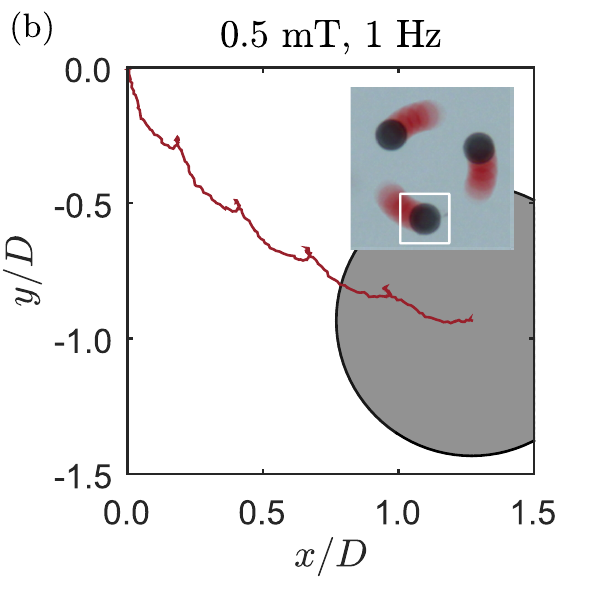}
	\refstepcounter{subfigure}
	\label{snaps:b}
\end{subfigure}
\begin{subfigure}{0.49\linewidth}
	\includegraphics[width=\linewidth]{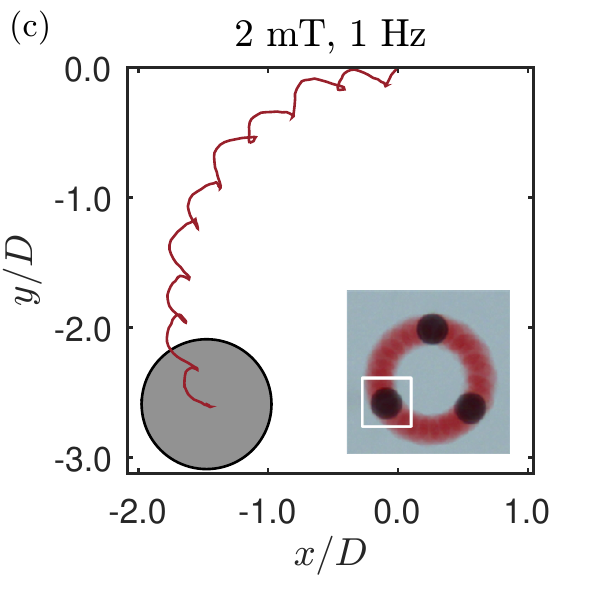}
	\refstepcounter{subfigure}
	\label{snaps:c}
\end{subfigure}
\begin{subfigure}{0.49\linewidth}
	\includegraphics[width=\linewidth]{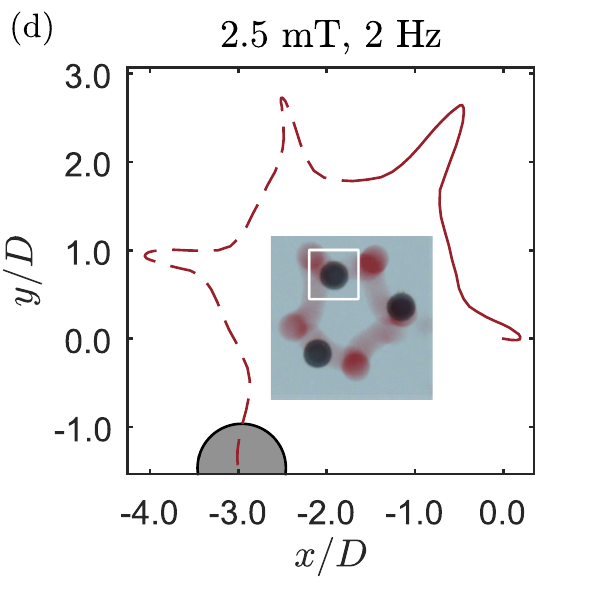}
	\refstepcounter{subfigure}
	\label{snaps:d}
\end{subfigure}
\caption{\textbf{Experimental trajectories.} --
Trajectories of a particle of dia\-meter $D = \SI{500}{\micro\meter}$ in an assembly exposed to a vertical field $B_z = \SI{4.9}{\milli\tesla}$ and a rotating horizontal field, for different values of amplitude $B_h$ and frequency $f$.
In red, the trajectories are shown for 5 periods.
The dark circle shows the final position of the particle and its size.
The insets show the trajectory for the whole assembly, with the corresponding particle in a white box.
In the bottom right case, the inset shows only the first second, corresponding to the solid line in the plot.}
\label{snaps}
\end{figure}

The role of $B_z$ is to oppose this capillary attraction with a dipole-dipole magnetic repulsion.
The pair potential for two particles of magnetic moment $\vec{m}$ is
\begin{equation}
U_m = -\frac{\mu_0}{4\pi d^3} \left( 3 (\vec{m}\cdot\vec{e})(\vec{m}\cdot\vec{e}) - \vec{m}\cdot\vec{m} \right).
\end{equation} 
If we neglect the contribution of the permanent magnetization $\vec{m}_0$, this can be rewritten as
\begin{equation} 
U_m = \frac{\mu_0\,m^2_z}{4\pi d^3} + \frac{\mu_0\,m^2_h}{4\pi d^3} \left(1-3\cos^2\phi\right),
\label{Um}
\end{equation} 
where $m_z$ and $m_h$ represent the vertical and horizontal components of $\vec{m}$ and $\phi$ is the angle between the pair of particles and $\vec{B}_h$.
This means that the contribution of the vertical field $\vec{B}_z$ is an isotropic repulsion.

The competition of the lateral capillary force and this dipole-dipole repulsion can lead to an equilibrium distance between particles.
In the absence of a horizontal field, three particles assemble into an equilateral triangle~\cite{vandewalle2012}.
The addition of a horizontal field $\vec{B}_h$ can break the symmetry of the assembly, which deforms into an isosceles under a quasi-static increase of $B_h$~\cite{grosjean2015}.

\section{Dynamics of a triangular assembly}

The angular frequency of the field $\omega = 2\pi f$ is chosen to be positive, corresponding to a counterclockwise rotation.
If we denote the average rotation speed of the assembly $\Omega$, we have $0\leq\Omega\leq\omega$.
\Cref{snaps} shows the overall motion of the assembly over 5 periods of the rotating field, as well as the trajectory of one of the three particles in more detail.
These four trajectories are also visible in video format, as part of the online supplementary information accompanying this article.
First, \cref{snaps:a,snaps:b,snaps:c} show the trajectories for three increasing values of the horizontal amplitude $B_h$ with the same applied frequency $f = \SI{1}{\hertz}$.
It can be seen that an increase in $B_h$ leads to an increase in the rotation speed.
The trajectory of each particle resembles an epicycloid, globally following a circle but punctuated by periodical cusps corresponding to each passing of the field.
The instantaneous rotation speed can briefly dip below zero during these cusps.
For larger values of $B_h$, the frequency of the cusps is doubled, suggesting a departure from the low-amplitude, linear regime that is discussed later on.

\begin{figure}
\includegraphics[width=.49\linewidth]{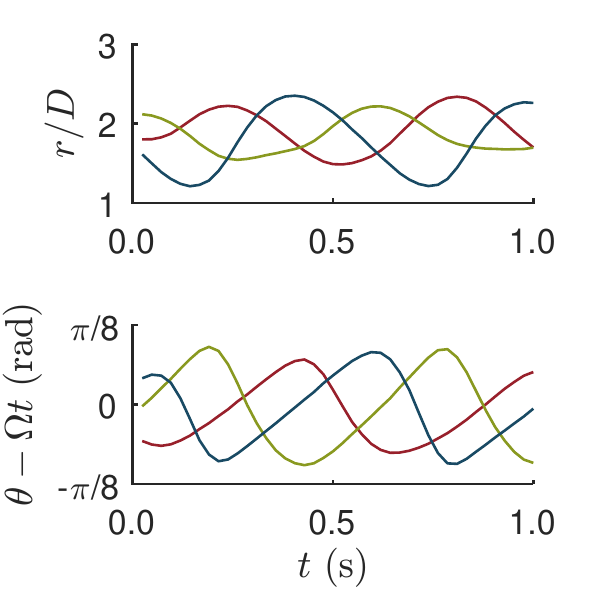}
\includegraphics[width=.49\linewidth]{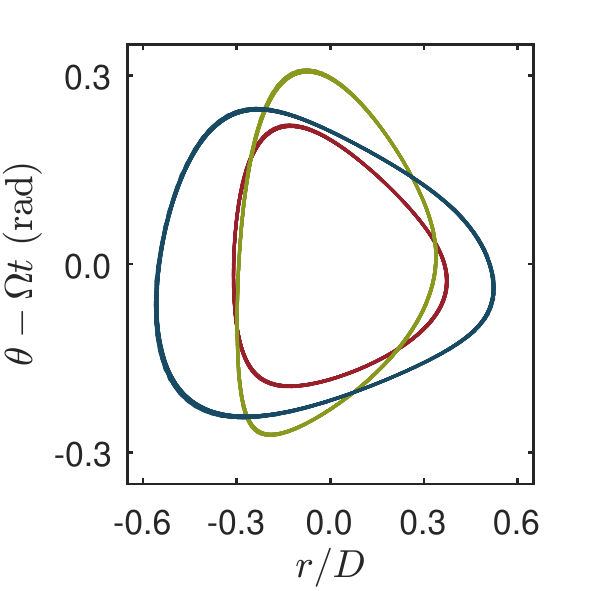}
\caption{\textbf{The high-frequency, high-amplitude regime.} --
On the left, positions in polar coordinates of the three particles from \cref{snaps}~(d) over \SI{1}{\second}.
On the right, cycles in the plane defined by the polar coordinates over \SI{20}{\second}.
Each color corresponds to one particle.}
\label{juggler}
\end{figure}

\Cref{snaps:d} shows the trajectory of one particle for even higher values of $B_h$ and higher $f$.
The rotation speed $\Omega$ is dramatically increased in this regime, performing a full turn in only about 6 periods of oscillation.
Both the rotation speed and the amplitude of deformation dramatically increase.
The trajectory is also quantitatively different, as there is no retrograde motion anymore.
This regime is characterized by much larger deformations, somewhat resembling a juggling pattern.
Typically, the particles oscillate sinusoidally, each separated by a phase of $2\pi/3$, as seen in the temporal evolution of $r$ and $\theta$ in \cref{juggler}.
The radial and angular components are roughly in quadrature, so that the particles follow open cycles in the ($r$, $\theta$) plane.
Note that for legibility, $\theta-\Omega t$ is used instead of $\theta$, which shows the evolution of $\theta$ in the referential of the rotating assembly.
The cycles have been filtered using a Fourier transform to remove non-periodic parts of the signal.

Cycles in the configuration space are synonymous with non-reciprocal motion, a necessary condition for locomotion at low Reynolds number~\cite{purcell1977,lauga2009}.
The Reynolds number in the experiment spans several orders of magnitude depending on $f$ and $B_h$. For instance, we have $\mathrm{Re} \sim 10^{-2}$ in \cref{snaps:a} and up to $\mathrm{Re} \sim 1$ in the juggling regimes of \cref{snaps:d}.
The fact that the individual particles in \cref{juggler} each describe a cycle, with a phase difference between neighboring particles, might be reminiscent of the swimming mechanisms adopted by some living organisms. 
Ciliates are covered in slender organelles called cilia.
They each individually describe a cyclic trajectory, with a phase difference between neighboring cilia such that a wave is propagated along the body of the bacteria~\cite{lauga2009}.
Recently, theoretical studies of ciliated motion have relied on particles describing cyclic trajectories~\cite{maestro2018,meng2019}.
Using these principles, magnetocapillary self-assemblies could serve as a simpler, self-assembled alternative to microfabricated systems~\cite{belardi2011,den2013}.
With simple cilia composed of one particle each~\cite{vilfan2010,kokot2011} could be beneficial to improve our fundamental understanding of ciliated locomotion.
Future work could expand this principle to larger, many-particle systems.
However, fully characterizing the dynamics and mechanisms at play in the simpler, three-particle system is an essential first step.

\begin{figure}[t]
\includegraphics[width=\linewidth]{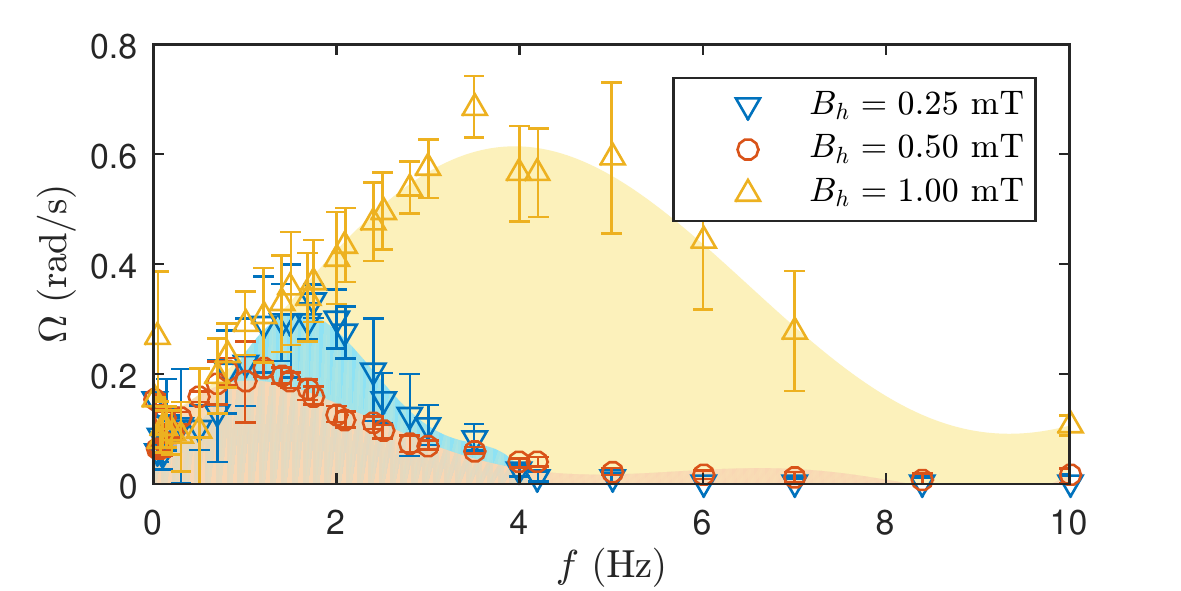}
\includegraphics[width=\linewidth]{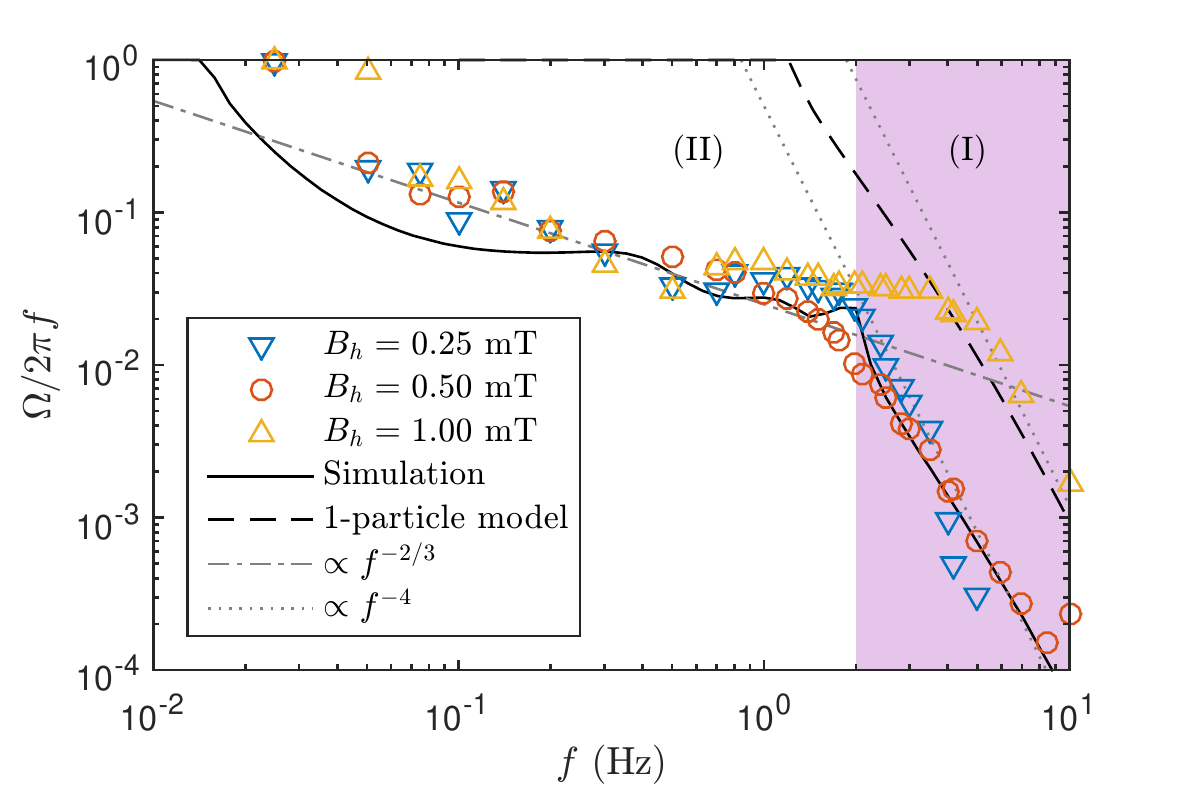}
\caption{\textbf{Influence of the frequency.} --
On top, the rotation speed of the assembly $\Omega$ reaches a maximum for a given value of the forcing frequency $f$. 
Below, dimensionless rotation speed $\Omega/2\pi f$ allows to see that, at very low $f$, the assembly follows the external field.
Two power laws are shown.
We have $\Omega/2\pi f \propto f^{-2/3}$ and $\Omega/2\pi f \propto f^{-4}$ on either side of the maximum.
This corresponds to regimes (I) and (II) in \cref{rotbx}.
Numerical simulations recover a similar behavior, but for a larger amplitude of \SI{2.2}{\milli\tesla}.
The 1-particle model from \cref{Eq:beadsRot} recovers regime (I) as well as the low-frequency locking.}
\label{rotfreq}
\end{figure}

\begin{figure}
\includegraphics[width=.49\linewidth]{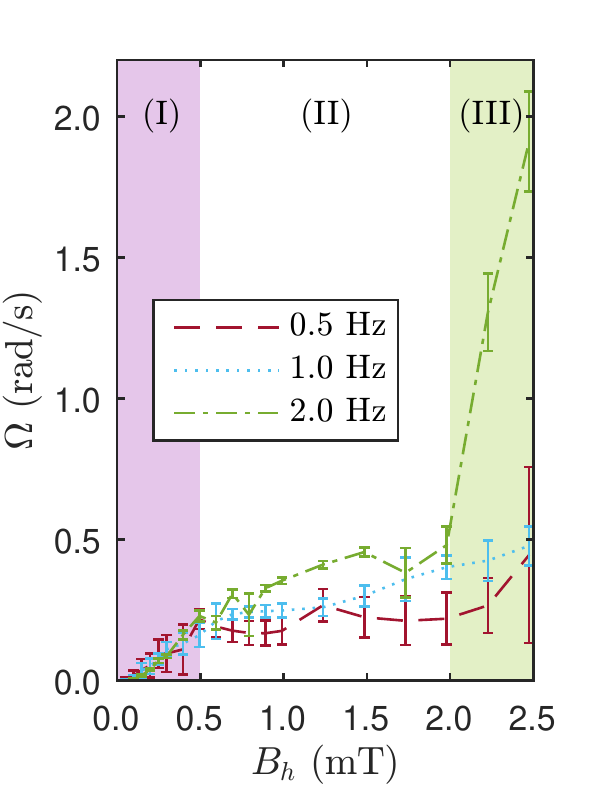}
\includegraphics[width=.49\linewidth]{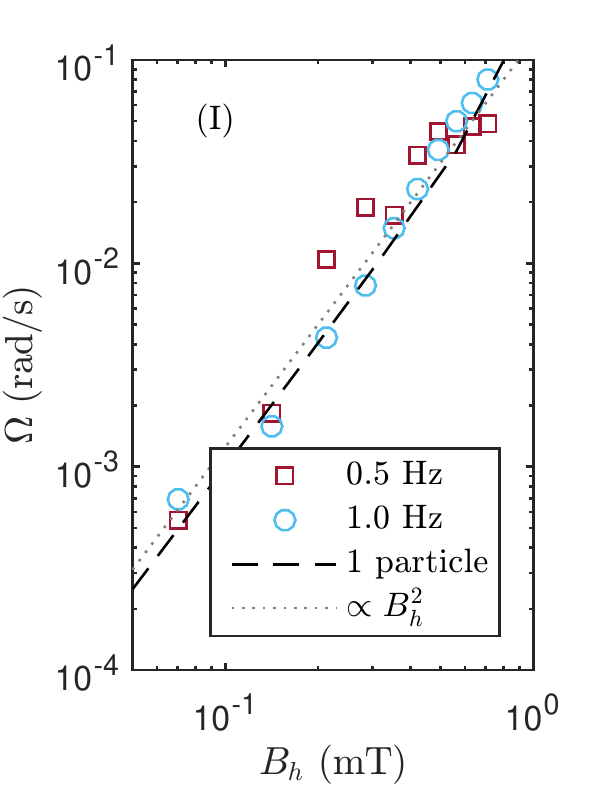}
\caption{\textbf{Influence of the amplitude.} --
On the left, rotation speed $\Omega$ as a function of the amplitude $B_h$, for three values of $f$.
Three regimes can be distinguished: the low-amplitude regime (I), the complex intermediate regime (II) and the `juggling' regime (III).
On the right, logarithmic plot of the low-amplitude regime.
The 1-particle model from \cref{Eq:beadsRot} has been adjusted on the data to show its quadratic behavior.}
\label{rotbx}
\end{figure}

\Cref{rotfreq,rotbx} show the influence of the two forcing parameters, respectively frequency $f$ and amplitude $B_h$, on the average rotation speed $\Omega$ of the system.
The value of $\Omega$ is measured from a linear regression over typically about 100 revolutions of the magnetic field.
Measurement errors are typically smaller than the symbols thanks to the high spatial and temporal resolution of the video acquisition.
When error bars are included, they correspond to the deviation from the linear regression.
This takes into account the oscillations of $\Omega$ over each passing of the field as well as slight variations of $\Omega$ that can occur over longer periods of time.

In \cref{rotfreq}, the frequency $f$ is varied for three values of the forcing amplitude.
One can see that the rotation speed reaches a maximum that is dependent on the forcing amplitude $B_h$, as evidenced in the linear plot of $\Omega$.
The bottom of \cref{rotfreq} shows the same experimental data in a logarithmic plot and with rotation speed $\Omega$ nondimensionalized by the forcing frequency.
This means that $\Omega/2\pi f = 1$ corresponds to the assembly simply following the external field.
This locking with the external field can only be observed for extremely low forcing frequencies, of the order of \SI{e-2}{\hertz}.
For larger values of $f$, $\Omega/2\pi f$ follows a broken power law with a transition that corresponds to a maximum of $\Omega$.
Dotted and dashed lines help to underline the power laws on each side of the maximum.
In terms of $\Omega$, we have an increase $\Omega \sim f^{1/3}$, followed by the maximum, then a decrease $\Omega \sim f^{-3}$.
The latter is discussed in \cref{model}.
The position of the maximum depends on the forcing amplitude $B_h$ in a non-trivial way.

\Cref{rotbx} shows the influence of the forcing amplitude $B_h$ on rotation speed $\Omega$, for three values of the forcing frequency $f$.
Three regimes have been distinguished.
At low forcing amplitude (I), $\Omega$ increases with $B_h$.
For the lowest values of the forcing frequency $f$, this increase is quadratic and independent on $f$, as shown by the logarithmic plot in \cref{rotbx}.
For intermediate amplitudes (II), $\Omega$ reaches a plateau.
This intermediate regime is complex and highly nonlinear, with local extrema and a stronger dependence in $f$.
At combined high amplitude and high frequency (III), $\Omega$ increases rapidly.
This corresponds to the `juggling' regime from \cref{juggler}.

\begin{figure}[t]
\centering
\includegraphics[width=\linewidth]{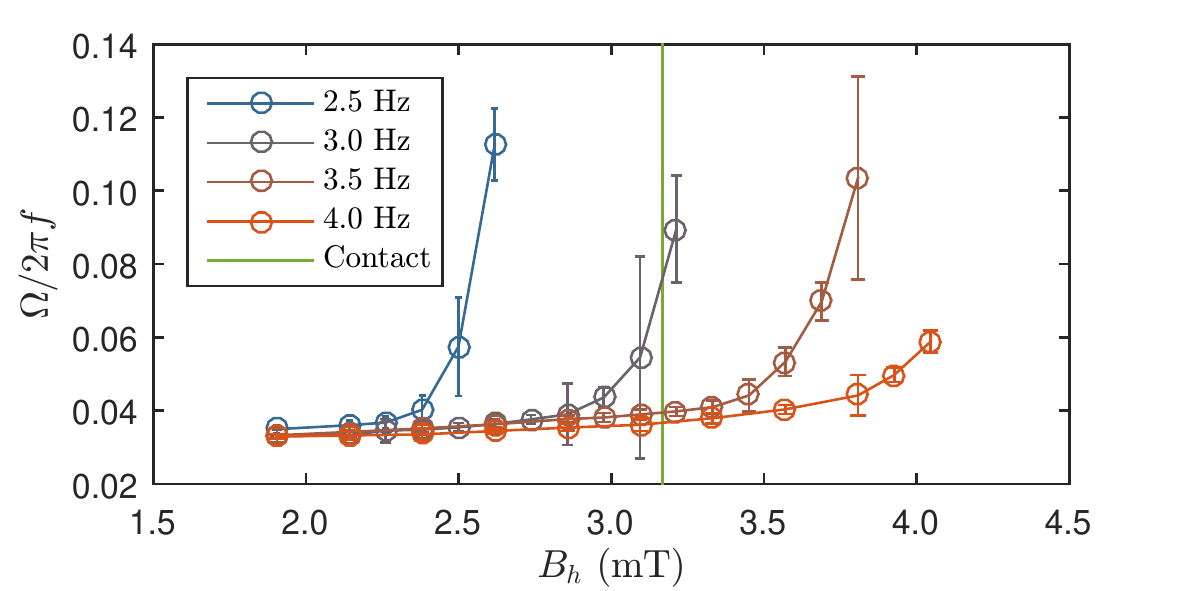}
\includegraphics[width=\linewidth]{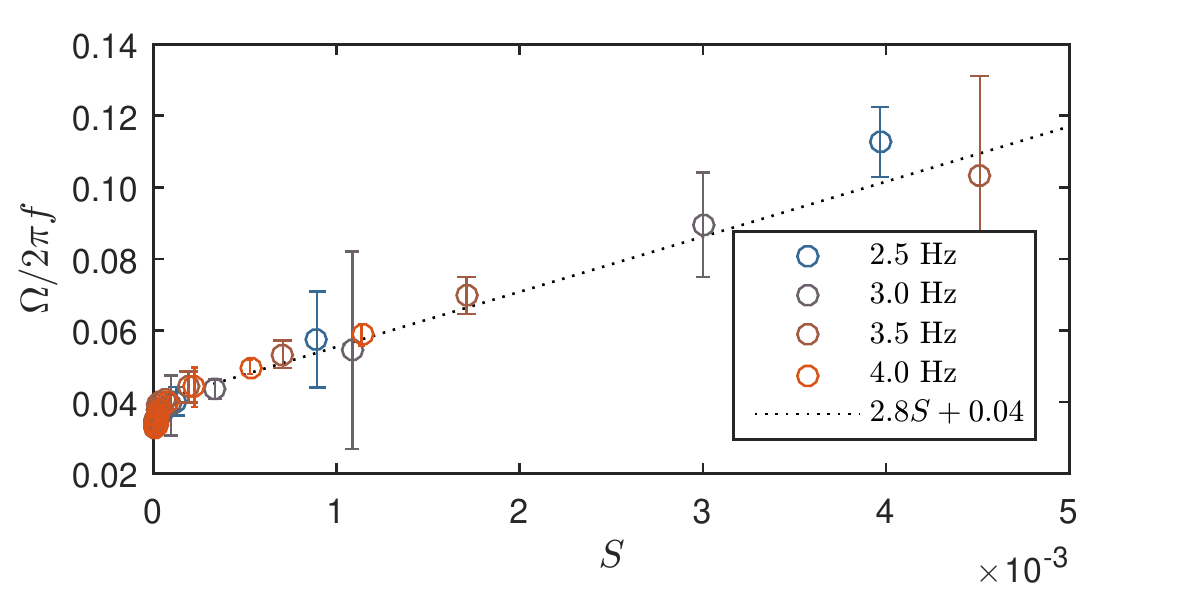}
\caption{\textbf{The juggling regime.} --
On top, $\Omega$ increases dramatically in the high-deformation regime, right before the particles come into contact.
The green line shows the quasi-static contact threshold as calculated in~\cite{chinomona2015}.
Below, $\Omega$ is shown as a function of the surface $S$ of the cycles in the $(d,\theta)$ plane.
The dotted line illustrates the linear relationship of \cref{ansatz2}.}
\label{area}
\end{figure}

This regime is discussed in more detail in \cref{area}, where the dimensionless rotational speed is measured as a function of $B_h$.
At lower amplitudes, we have roughly $\Omega/2\pi f \approx 0.2$.
Right before the particles come into contact, the speed drastically increases.
This is accompanied by much larger deformations like those shown in \cref{juggler}.
The theoretical value for the collapse of the assembly is also shown.
One can see that, for excitation frequencies larger than \SI{3}{\hertz}, the `juggling' mode can persist above the theoretical collapse.
Indeed, a static field $\vec{B}_{h,\;0} = B_{h,\;0} \;\vec{e}_x$ of more than \SI{3.2}{\milli\tesla} would cause the particles to come into contact~\cite{chinomona2015}.
This indicates that the particles are not at equilibrium, and that the inertia of the particles therefore plays a role.
Indeed, the Reynolds number in this case approaches unity.
The bottom of \cref{area} compares the rotation speed with the area $S$ of the cycles identified in \cref{juggler}.
The relation between $S$ and $\Omega$ is discussed in \cref{deformations}.

\section{Discussion}
\label{discussion}

Three mechanisms can potentially play a role in the rotation of the assembly.
First, if the particles are not perfectly isotropic, the rotation could be driven by the individual rotations of the particles.
Secondly, the whole structure could reorient in an external field due to the pairwise magnetic dipolar interactions between particles.
Thirdly, non-reciprocal deformations could power the rotation in a similar fashion to microswimmers.

\subsection{Individual rotations}
\label{particles}

To unravel the origin of the ferromagnetic contribution to the total magnetization of the particles discussed in \cref{Eq:TranslBeads}, we performed series of micromagnetic simulations which are based on the solution of the Landau-Lifshitz-Gilbert equation~\cite{landau1935,gilbert2004} for the magnetization motion for relevant internal and external magnetic energy contributions in a bulk.
We use the mumax3 simulation package optimized for graphics processing units (GPU)~\cite{vansteenkiste2014}.

Though modern computational power allows for reaching dozens of millions of interacting magnetic cells, the physical size of simulated systems is limited to several (2-3) micrometers.
At this scale we witness a rapid drop of the remanent magnetization (below one percent) of a single perfectly spherical bead mainly because of the formation of magnetic domains.
A further increase in size up to desired \SI{500}{\micro\meter} should result in even lower net magnetization.
In addition, thermal fluctuations of the magnetic order parameter are known to enhance the effect of the magnetization reduction.
Further effects like polycrystalline structure of the beads at such sizes should also lead to a vanishing remanent magnetization, as the average over all directions of polycrystalline axes in space should be zero. 
We conclude that the performed micromagnetic simulations suggest zero remanent magnetization for particles with diameters \SI{500}{\micro\meter} and for parameters related to bulk steel alloy.

Non-sphericity of the beads might also explain the existence of the permanent internal magnetic moment.
However, for particles used herein (precision ball bearings of AMBA grade 50), the deviation 
from a perfect sphere is at most \SI{1.2}{\micro\meter} or \SI{0.25}{\percent}.
Using known expressions of the shape anisotropy energy of an ellipsoid and assuming a monodomain particle~\cite{beleggia2006,coey2010}, we estimate that such a deviation in shape would only induce a magnetization of about \SI{0.1}{\percent} of the saturation one.
A more plausible explanation for the magnetic anisotropy is that the internal structure of the particles might be affected by the manufacturing process.
Indeed, the spheres are formed by cutting the steel wire into small cylinders, which are then pressed into a spherical die and finally rounded.
The process of pressing the cylinders into a die (heading) might create an anisotropy in the spheres, which could be at the origin the permanent component of the magnetization.

\begin{figure}[t]
\centering
\includegraphics[width=\linewidth]{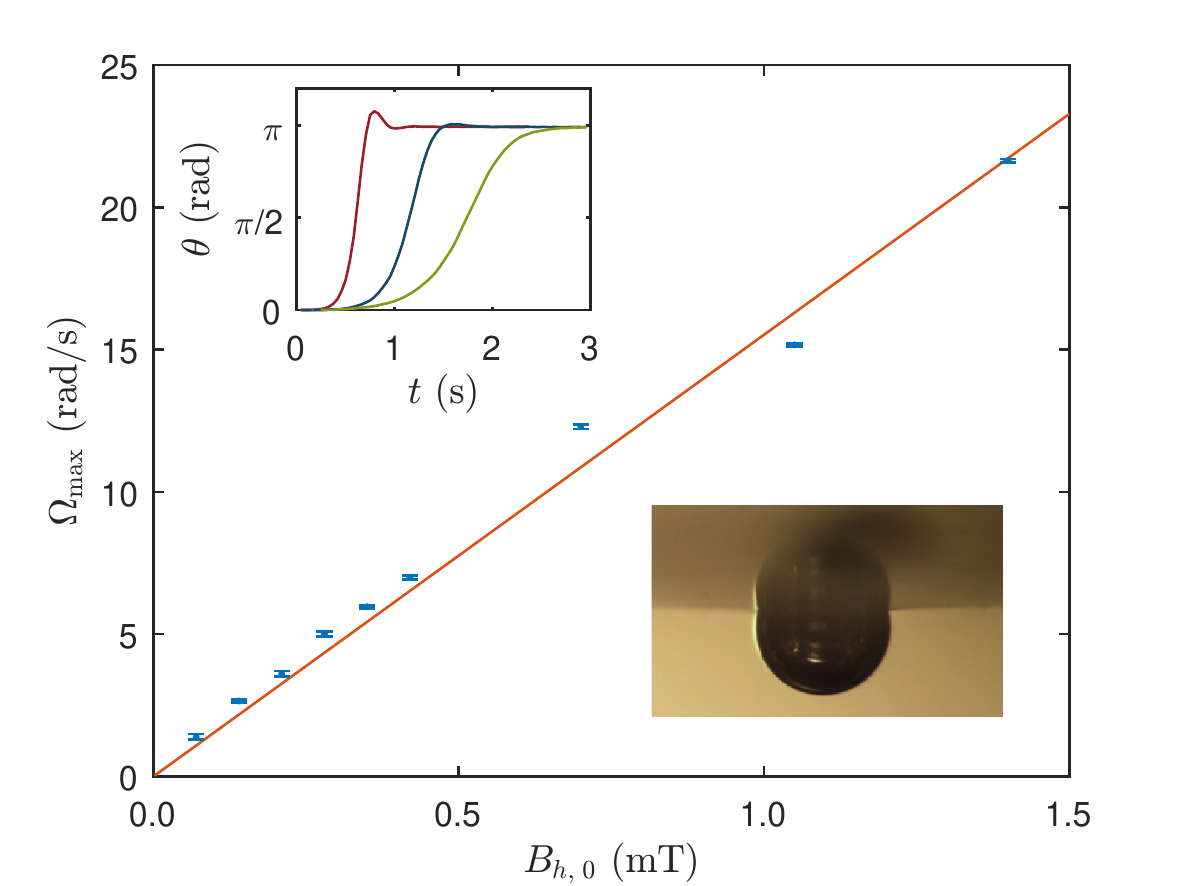}
\caption{\textbf{Single particle.} --
Maximal rotation speed $\Omega_{\mathrm{max}}$ of a single particle, as a function of the amplitude of a horizontal field $\vec{B}_{h,\;0}$ that changes sign at $t=0$.
The error bars correspond to the uncertainty from the correlation.
The top left inset shows the angle $\theta$ as a function of time for three values of $B_{h,\;0}$.
The bottom right inset shows a side view of a particle from below the interface.}
\label{1bead}
\end{figure}

This permanent component is essential to account for the individual rotation of the particles.
Evidence of such rotation can be found by tracing a spot on a particle with a marker pen.
By correlation, we can then calculate the angle $\theta$ of the bead at any time.
If we consider that we are in the Stokes regime, then the magnetic torque 
\begin{equation}
\tau_m = |\vec{m} \times \vec{B}|,
\label{magnetictorque}
\end{equation}
where $\vec{m}$ is the magnetic moment of the particle, has to be equal at all times to the viscous torque, which for a fully immersed sphere rotating at a constant speed $\Omega$ reads
\begin{equation}
\tau_\eta = 8 \pi \eta R^3 \Omega.
\label{viscoustorque}
\end{equation}
We can therefore expect a direct proportionality between $\tau_m$ and $\Omega$.
In \cref{1bead}, a single sphere is exposed to a horizontal constant field $B_{h,\;0}$ that changes sign abruptly, leading the particle to rotate a half turn.
The maximal rotation speed $\Omega_{\mathrm{max}}$ is measured for several values of $B_{h,\;0}$.
Two possible evolutions of $\Omega_{\mathrm{max}}$ with $B_{h,\;0}$ can be expected.
If the magnetic moment $\vec{m}$ responsible for the rotation is permanent and of magnitude $m_{\mathrm{perm}}$, we have
\begin{equation}
\Omega_{\mathrm{max}} \propto \tau_{m,\ \mathrm{max}} = m_{\mathrm{perm}} B_{h,\;0} \propto B_{h,\;0}.
\label{permanent}
\end{equation}
On the other hand, if the moment responsible is an induced moment of magnitude $m_{\mathrm{ind}} = \chi V B_{h,\;0} / \mu_0$, then we have
\begin{equation}
\Omega_{\mathrm{max}} \propto \tau_{m,\ \mathrm{max}} = m_{\mathrm{ind}} B_{h,\;0} \propto B_{h,\;0}^2.
\label{induced}
\end{equation}
As can be seen in \cref{1bead}, the evolution of $\Omega_{\mathrm{max}}$ with regards to $B_{h,\;0}$ is linear, suggesting that the rotation of the individual particles is indeed an effect of a small remanent magnetization.

The top left inset of \cref{1bead} shows the angle $\theta$ as a function of time during the inversion.
One can see that a stronger field leads to a faster rotation.
For higher values of $B_{h,\;0}$, there is a slight overshoot due to inertia.
This is not surprising considering that, in order to facilitate the measurement, a larger particle of \SI{1}{\milli\meter} in diameter was used.
It was observed that the axis of rotation is always perpendicular to the interface.
This might be surprising considering that the direction of $\vec{m}_0$ depends on the random orientation of the particle when dropped.
However, even the strongest horizontal fields used herein, about \SI{50}{\milli\tesla}, fail to produce any measurable out-of-plane rotation.
These types of rotations might be prevented due to the additional energy cost required to move the contact line~\cite{sukhov2019}.
The bottom right inset of \cref{1bead} shows a side view of a particle.
The contact angle is approximately \SI{80}{\degree}.
The contact line is generally axisymmetric, though its position on the particle can vary depending on how the particle was placed at the interface.

These individual rotations likely power the rotational motion at the lower excitation amplitudes, where the effect of the permanent component of the magnetization dominates.
This corresponds to regime~(I) in \cref{rotbx}.
As the amplitude increases, the effect of the induced component of the magnetization becomes important through dipole-dipole interactions.

\subsection{Dipole-dipole alignment}

Let us consider only induced magnetic dipoles, which are always aligned with the external field.
From \cref{Um}, the potential energy associated with the horizontal field for a pair of particles can be written as
\begin{equation}
U_{m,\;x}^{\mathrm{pair}} = \frac{\mu_0\,m^2}{4\pi d^3} \left(1-3\cos^2\phi\right) = U_0 \left(1-3\cos^2\phi\right),
\end{equation}
where $U_0$ corresponds to the maximum energy for a given $d$, reached when the pair is oriented perpendicular to the field.
The minimum of energy is reached when the pair is aligned with the field and is equal to $-2 \,U_0$.
This means that under a horizontal field $\vec{B}_{h,\;0}$ a pair of particles experiences a torque and tends to align with the field.

Let us now consider the case of three particles.
For simplicity and because this situation is prevalent in the experiment~\cite{grosjean2015,grosjean2018}, we suppose that the particles lay on the vertices of an isosceles triangle.
Let $\alpha$ denote the vertex angle of the isosceles and $\phi$ the angle between its base and the external field.
Summing the pairwise energy contributions leads to
\begin{equation}
U_{m,\;x}^{\mathrm{iso}} = U_0 \left(3\cos\alpha\cos 2\phi - 1 + \frac{1-3 \cos^2 \phi}{8 \sin^3 \frac{\alpha}{2}} \right).
\label{potentialiso}
\end{equation}
For two values of $\alpha$ between 0 and $\pi$, this expression is independent on $\phi$, meaning that all the orientations of the assembly with regard to the field are energetically equivalent.
These values are (i) $\alpha = \pi/3$, corresponding to an equilateral triangle, and (ii) $\alpha \approx 1.2605$, which is the value of $\alpha$ such that $\cos\alpha\sin^3(\alpha/2) = 1/16$.
In fact, by symmetry, the independence of the total magnetic interaction energy on the orientation of a rotating field $\vec{B}_h$ is verified for any regular polygon.
This means that, for low values of $B_h$, the dynamics is driven by the individual rotations of the particles for which the torque is proportional to $B_h$.
Conversely, we can expect the dipole-dipole interaction, quadratic in $B_h$, to play a role at high amplitudes, for which the deformation of the triangle is significant.

Considering that both individual rotations and deformations of the structure can be observed in regime~(II), the complex dynamics observed in this intermediate regime might be due to a combination of the two effects.
Both the induced and the permanent components of the magnetization expressed in \cref{Eq:TranslBeads} compete, though it is difficult to isolate the influence of each effect experimentally.
Other effects might add to the complexity of the intermediate regime.
For instance, self-assemblies of rotating particles have been shown to form due to hydrodynamic repulsions~\cite{grzybowski2000}, though the typical magnitude of this force $\mathcal{F} \sim \rho \omega^2 a^7 / d^3$ is orders of magnitude smaller in our case.
At even higher amplitudes, we can expect the contribution of $\vec{m}_0$ to become negligible.
As we will show next, the dipole-dipole alignment mechanism also becomes ineffective at higher frequencies, such as in regime~(III).
A third mechanism is therefore needed to explain this high-deformation regime.

\subsection{Rotational locomotion}
\label{deformations}

Under a vertical magnetic field, magnetocapillary self-assemblies often adopt the shape of a regular polygon~\cite{lumay2013}, with notable exceptions at $N=2$ and $N=5$.
These regular assemblies only have a preferred orientation when the horizontal field is strong enough to significantly alter their shape.
However, both this effect and the effect of the individual rotations of the particles should become inefficient at high frequencies, as the dipoles struggle to follow the external field.
For instance, equating the magnetic torque from \cref{magnetictorque} to the viscous torque from \cref{viscoustorque} and neglecting inertia, we find
\begin{equation}
\sin\theta = \frac{8\pi\eta R^3}{m B_h} \dot{\theta} = \tau \dot{\theta},
\label{compass}
\end{equation}
which has a solution of the type
\begin{equation}
\theta = 2 \cot^{-1} \left( e^{-t/\tau+C_0} \right),
\label{solution}
\end{equation}
where $C_0$ is an integration constant.
This means that there is a typical reaction time $\tau$ needed for a dipole to align with an external field.
In fact, looking at \cref{solution}, we can see that it takes about $5\tau$ to cover \SI{95}{\percent} of the half-turn.
We can therefore define a typical cut-off frequency
\begin{equation}
f_{\mathrm{cut}} = \frac{1}{\tau} = \frac{m B_h}{8\pi\eta R^3}.
\label{cutoff}
\end{equation}
which is equal to \SI{10}{\hertz} for a field of about \SI{0.6}{\milli\tesla}.
This would mean that there is an observable decrease in the oscillation amplitude as early as \SI{2}{\hertz}.
Dipole-dipole interactions are usually weaker than dipole-field interactions in magnetocapillary assemblies, as the field generated from a particle is given by $\mu_0 m/4\pi d^3$, which for an induced dipole is smaller by a factor $R^3/d^3 \approx 1/64$ for particles at a typical distance $2D$.
This means that we can expect an even more drastic cut-off in the case of the dipole-dipole alignment.
This is consistent with the extremely low excitation frequency needed to observe a rotation of the whole assembly locked with the external field, as shown in \cref{rotfreq}.

This means that both previously discussed mechanisms cannot explain the sudden increase in rotational speed at combined high frequency and high amplitude described in \cref{juggler}.
As deformations are high in this regime, one might wonder if the motion is powered by non-reciprocal deformations.
In other words, the cycles shown in \cref{juggler} could power the rotational motion, similarly to the translational motion of the Najafi-Golestanian swimmer~\cite{golestanian2008}.
Let us assume that each particle, represented by its polar coordinates $d_i/D$ and $\theta_i$, oscillates harmonically both radially and angularly at a frequency $f$.
We denote the corresponding amplitudes $A_{d,i}$ and $A_{\theta,i}$ and the phase between both signals $\phi_i$.
By analogy with~\cite{golestanian2008}, we construct the ansatz of an expression for the rotational speed
\begin{equation}
\Omega \propto \sum_{i=1}^{N} 2\pi f \; A_{d,\;i} \; A_{\theta,\;i} \sin \left( \phi_i \right),
\label{ansatz}
\end{equation}
where we consider each particle to act as an independent motor whose contributions can be summed.
As the trajectories in the $(d,\theta)$ plane are not necessarily ellipses, we can assume a more general expression
\begin{equation}
\Omega \propto \sum_{i=1}^{N} 2\pi f S_i = 2\pi f S,
\label{ansatz2}
\end{equation}
where $S_i$ is the dimensionless surface enclosed by the trajectory of particle $i$.
The expressions from \cref{ansatz} and \cref{ansatz2} are equivalent in the case of an ellipse.
\Cref{area} compares the rotational speed and the dimensionless surface of the cycles for the data of \cref{area}.
Most of the points are concentrated around $\Omega/2\pi f \approx 0.04$ and $S \approx 0$, corresponding to the plateau in \cref{area}, where deformations are small and the rotation is explained by the two previously discussed mechanisms.
However, for fast rotations $\Omega/2\pi f > 0.04$ we see a linear relationship between the combined dimensionless area $S$ and the dimensionless rotational speed $\Omega/2\pi f$, confirming our hypothesis.

\section{Model}
\label{model}

In order to address the question of the importance of each dynamics described in \cref{discussion}, we develop a numerical model of the experiments presented in this article.
The algorithm integrates Newton's equations for each bead. 
As the beads both translate and rotate on the surface, an inventory of all the forces and torques is required.

To describe the translation, one has to take into account the Cheerios effect introduced in \cref{Uc}, the magnetic dipole-dipole interaction introduced in \cref{Um} as well as the hydrodynamic interactions. 
The translation of a bead generates a stokeslet, \emph{i.e.} the flow field originating from a point force~\cite{blake1974}.
This flow drags neighboring beads with the following force 
\begin{equation}
\vec{F}_{h,\;i}^t = \sum_{j\neq i}\left( \frac{9\pi\eta R^2}{2d}\left(\vec{v}_j + \left(\vec{e}.\vec{v}_j\right)\vec{e}\right) + \frac{3\pi\eta R^4}{2d^3}\left(\vec{v}_j - 3\left(\vec{e}.\vec{v}_j\right)\vec{e}\right)      \right),
\end{equation}
where the labels $i$ and $j$ refer to two beads on the surface and $\vec{v}$ denotes their speed. 
To express the forces as a function of the translation and rotation speeds, the substitutions $v_i \rightarrow f_i/6\pi\eta R$ and $f_j \rightarrow 6\pi\eta R v_j$, which are typically valid up to $\mathcal{O}(d^{-4})$, have been applied~\cite{mazur1982}.
Since the beads rotate in the plane of the interface, the flow field generated by a point torque, a rotlet, must also be taken into account~\cite{blake1974}.
Using the same substitution~\cite{mazur1982}, we find
\begin{equation}
\vec{F}_{h,\;i}^r = \sum_{j\neq i}\frac{6\pi\eta R^4}{d^3}\left(\vec{d}\times\vec{\Omega}_j\right).
\end{equation}
For simplicity, we considered fully immersed spheres.
Previous investigations showed that the effect of the partial immersion is essentially to reduce the Stokes drag by a prefactor which was determined to be about $0.86$ for the particles used herein~\cite{lagubeau2016}.
A theoretical prediction for this factor can also be obtained from~\cite{dorr2016}, yielding a prefactor of $0.55$ for a contact angle of \SI{80}{\degree}.
This model assumes a perfectly flat interface and does not take the meniscus into account, which might explain the difference.

The forces associated with the Cheerios effect and the dipole-dipole interaction are obtained by taking the derivative of the corresponding potentials, \emph{i.e.} \cref{Uc,Um}.
Therefore the equations of motion are
\begin{equation}
	m_i\vec{a}_i = -\vec\nabla U_c -\vec\nabla U_m -6\pi\eta R\vec{v}_i +  \vec{F}_{h,\;i}^t +\vec{F}_{h,\;i}^r .
\label{Eq:newton}
\end{equation}
In this model, only pair interactions have been considered therefore justifying the expansion of the hydrodynamic coupling to the third order in $1/d$.

\begin{figure}[t]
\begin{subfigure}{0.49\linewidth}
	\includegraphics[width=\linewidth]{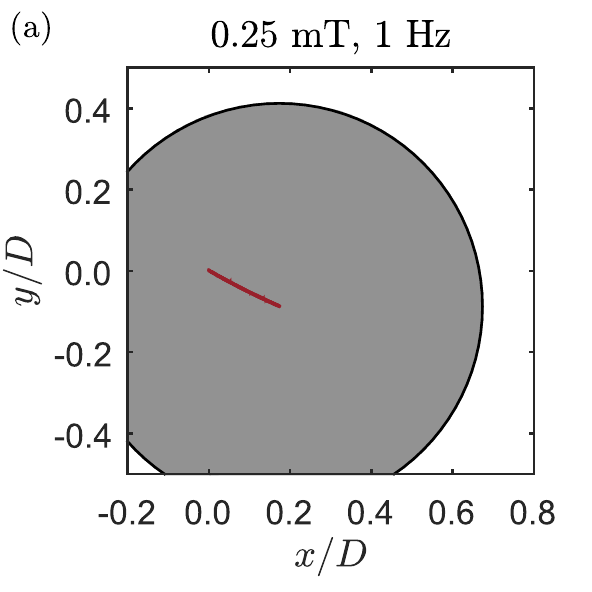}
	\refstepcounter{subfigure}
	\label{simusnaps:a}
\end{subfigure}
\begin{subfigure}{0.49\linewidth}
	\includegraphics[width=\linewidth]{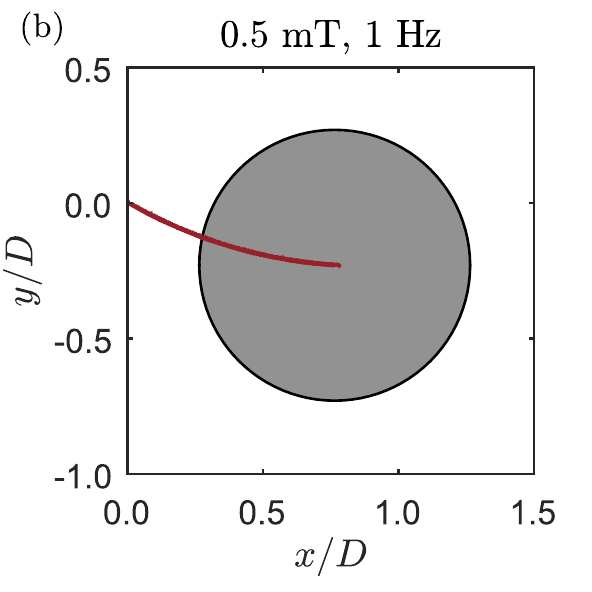}
	\refstepcounter{subfigure}
	\label{simusnaps:b}
\end{subfigure}
\begin{subfigure}{0.49\linewidth}
	\includegraphics[width=\linewidth]{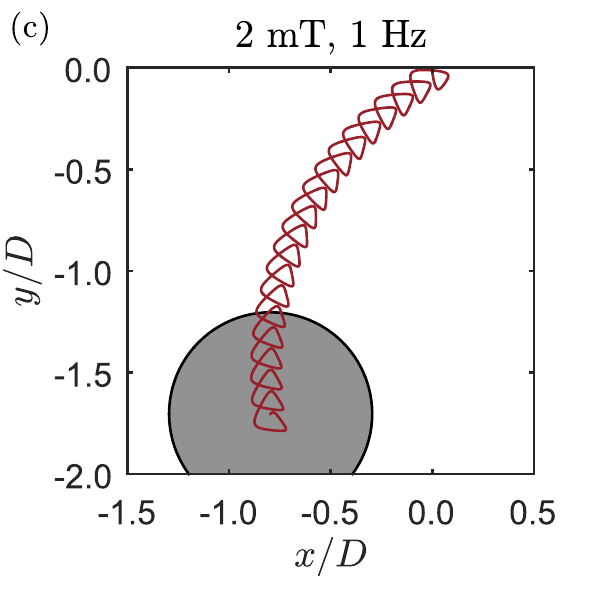}
	\refstepcounter{subfigure}
	\label{simusnaps:c}
\end{subfigure}
\begin{subfigure}{0.49\linewidth}
	\includegraphics[width=\linewidth]{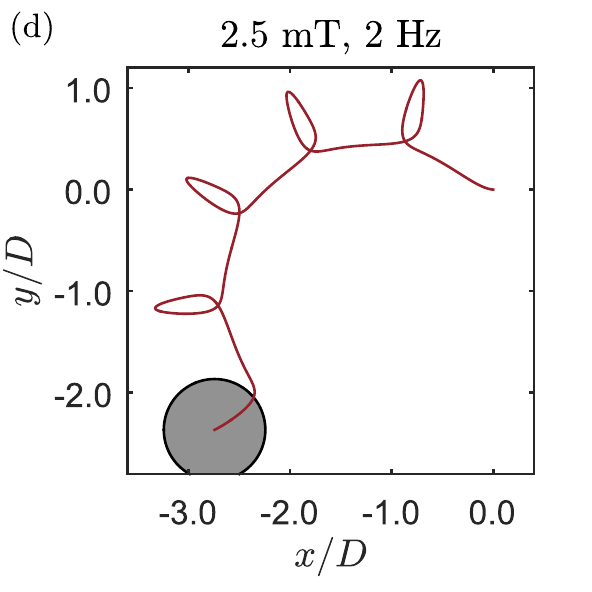}
	\refstepcounter{subfigure}
	\label{simusnaps:d}
\end{subfigure}
\caption{\textbf{Numerical trajectories.} --
Trajectory of a particle in a triangular assembly simulated using \cref{Eq:newton}.
The parameters are the same as in \cref{snaps}.
The trajectories are shown for 100, 100, 20 and 5 periods, respectively.}
\label{simusnaps}
\end{figure}

As stated previously the rotation of the beads comes from the magnetic torque arising from the residual magnetization. 
No torque can arise from the Cheerios effect under the assumption that the beads create a symmetric wetting line when dropped on the liquid interface. 
To describe the magnetic torque, we need to take into account the anisotropy in magnetic properties.
Using the ansatz of \cref{Eq:TranslBeads}, the rotation of the beads is described via the equation
\begin{equation}
	I\vec{\alpha}_i = -\vec{m}_{0,\;i}\times\vec{B} - 8\pi\eta R^3\vec{\Omega}_i,
	\label{Eq:RotBeads}
\end{equation}
where $I$ is the moment of inertia and $\vec{\alpha}$ the angular acceleration. 
This equation is projected along the $z$ direction as the rotation of each bead lies in the plane defined by the surface.
The equations of motion presented above are integrated numerically.
Note that each parameter appearing in those equations can be measured experimentally, leaving no free fitting parameter.
In order to illustrate the results provided by the numerics, \cref{simusnaps} gives the dynamics of the swimmer for the same parameters as in \cref{snaps}.
The dynamics obtained in simulations is essentially similar to the what is measured experimentally.
For small amplitude and frequency, the deformation of the triangle is small and the rotation slow.
When the amplitude is increased, the rotation becomes faster while the deformation of the structure becomes more important.
Finally, the nonlinear juggling regime for high amplitude and frequency is also observed.

Some differences with the experiment can also be noted.
First, a more pronounced retrograde motion is observed, as seen in \cref{simusnaps:c,simusnaps:d} when compared with \cref{snaps:c,snaps:d}.
Recovering these finer structures accurately might require a more precise description of the various terms in \cref{Eq:newton}, including higher-order hydrodynamic terms and taking into account the partial immersion of the spheres.
Secondly, the rotation speed is typically much lower than in the experiment, a notable exception being the high-deformation regime.
Taking into account the inertia of the flow might increase the numerical prediction for $\Omega$.
This explanation is insufficient, though, considering that the model is the most accurate when the Reynolds number is the highest.
The solution might reside in the description of the magnetic properties of the particles, as it becomes more complex at low amplitude.
The anisotropy of the particles' properties comes into play, and a perfectly soft-ferromagnetic description becomes insufficient to describe what is observed.
While the ansatz of \cref{Eq:TranslBeads} allows to recover the general behavior of the particles, the model would certainly benefit from a finer, three-dimensional study of their magnetic properties.
A better understanding of the magnetic properties of the particles could also improve reproducibility.
Indeed, the orientation of the anisotropy, modelled by $\vec{m}_0$, is not controlled.
This can lead to variations in the response of a particle to the field, depending on how it was placed at the interface.
Combined with the variations in contact line radius, this effect is likely the main cause of variations between experimental runs.

The numerical model allows to turn off some effects to determine their influence.
As expected, removing $\vec{m}_0$ leads to no detectable rotation in the low-amplitude regime of \cref{simusnaps:a,simusnaps:b}.
In the intermediate regime of \cref{simusnaps:c}, the overall speed is reduced but the motion qualitatively similar, showing that both the effect of $\vec{m}_0$ and the dipole-dipole interaction matter.
Finally, the high-amplitude, high-frequency regime is almost not affected by the removal of $\vec{m}_0$, which confirms that the anisotropy of the particles is not at play there.

This model also gives the possibility to describe the power laws observed in \cref{rotfreq,rotbx}. 
Let us assume that the triangle remains equilateral once exposed to the rotating magnetic field with one bead at each vertex.
This hypothesis corresponds to the observation in the linear regime.
As the beads rotate due to the external magnetic field, each of them is the source of a rotlet at the surface of the fluid.
Since the Stokes equation is linear, the global flow field is a rotlet centered at the geometrical center of the assembly.
This feature can be observed in \cref{ylo} by following the dyes around each bead.
The triangle therefore rotates with this induced flow.
Interestingly, again thanks to the linearity of the Stokes equation, the speed of rotation of the triangle is proportional to the speed of rotation of each bead.
Considering that each particle rotates at the same rate $\Omega_i$ and using the definition of a rotlet~\cite{blake1974}, one can find that 
\begin{equation}
\Omega = 2\Omega_i R^3/d^3.
\label{Eq:omegalow}
\end{equation}
As a consequence, quantifying the rotation of one bead allows to understand the global dynamics and to extract scaling which can be compared to experiments.
\Cref{Eq:RotBeads} can be written as
\begin{equation}
	I\alpha_i = - m_{0,\;i} B_h \sin\left(\theta_i-\omega t\right) - 8\pi\eta R^3 \Omega_i.
	\label{Eq:beadsRotdim}
\end{equation}
Introducing the dimensionless time $u = \omega t$ and the corresponding time derivative with dotted symbols, one has
\begin{equation}
	\ddot{\theta}_i + \frac{8\pi\eta R^3}{I\omega}\dot{\theta}_i + \frac{m_{0,\;i} B_h}{I\omega^2} \sin\left(\theta_i - u \right) =0.
	\label{Eq:beadsRot}
\end{equation}
This equation gives two dimensionless parameters $c_1 = 8\pi\eta R^3/I\omega$ and $c_2 = m_0 B_h/I\omega^2$, each depending on the frequency of oscillation of the magnetic field.
The dashed line in \cref{rotfreq} gives the average rotation speed $\Omega_i$ of a single bead as a function of the frequency $f$, using typical experimental parameters.
Two regimes are recovered.
Below \SI{1}{\hertz}, the particle is locked with the external field so that $\Omega_i/2\pi f=1$.
At higher frequency, we have roughly $\Omega_i \propto f^{-3}$, as the particle struggles to follow the field.
Compared to the three-particle assembly, the rotation is generally faster and the uncoupling with the rotating field happens at a much higher frequency.
The intermediate regime is absent, confirming that it is likely influenced by the dipole-dipole interactions in the assembly.
In \cref{rotbx}, $\Omega_i$ is shown as a function of $B_h$, which is proportional to $c_2$ as $c_1$ is kept constant. 
The curve has been adjusted on the experimental points, showing a very similar quadratic dependence in $B_h$. 
This means that in the low-amplitude linear regime, the assembly as a whole essentially behaves like a single effective dipole.

In general, it is not trivial to link the rotation of each particle $\Omega_i$ to the rotation of the assembly $\Omega$, as including inertia is necessary to reproduce the experimental observations in the higher-frequency regimes.
Indeed, if we neglect the left-hand side of \cref{Eq:beadsRotdim} and combine it with \cref{Eq:omegalow}, we find $\Omega \propto f$ at low frequency and $\Omega \propto f^{-1}$ at high frequency.
By taking the inertia of the particles into account, we find a third regime where $\Omega \propto f^{-3}$, which corresponds to what is observed experimentally at high frequency in \cref{rotfreq}.
The single-particle study from \cref{1bead} also shows a visible overshoot due to inertia, although this study was performed on a larger particle of \SI{1}{\milli\meter} in diameter.

Nonetheless, inertia in \cref{Eq:beadsRotdim} must be distinguished from the inertia in the flow.
If we compare the relative importance of inertia over viscous damping in the flow, we find the Reynolds number defined as $Re = \rho_f U L/\eta$.
Using $L \sim R$ and $U \sim R\omega$, one has 
\begin{equation}
	\mathrm{Re} \sim \frac{\rho_f R^2 \omega}{\eta}.
\end{equation}
Conversely, in \cref{Eq:beadsRotdim}, we can compare the inertial term $I\alpha$ with the damping one $8\pi\eta R^3 \omega$.
Using $\alpha \sim \omega^2$, one finds the dimensionless number
\begin{equation}
	\xi = \frac{I\alpha}{8\pi\eta R^3\eta} \sim \frac{\rho_s R^2 \omega}{\eta}.
\end{equation}
This means that depending on the density of the fluid $\rho_f$ and the density of the particle $\rho_s$, it is possible to remain in a viscosity-dominated flow while having a non-negligible influence of inertia on the bead's rotation.
In the experiment, we have $\rho_s/\rho_f \approx 7.8$.
Considering that $\mathrm{Re}$ can approach $1$ from below at the highest amplitudes and frequencies used herein, this density ratio means that we often have $\mathrm{Re}<1$ but $\xi>1$.
Compared to the Stokes regime, this intermediate region where both inertia and viscosity matter is seldom discussed in the literature.
Further studies on this aspect using magnetocapillary assemblies could therefore be beneficial to our general understanding of small-scale flows.

Particles of a few tens of microns in diameter could in theory still be bound by the magnetocapillary interaction~\cite{lagubeau2016}, assuming that the magnetic properties of the particles scale properly.
Indeed, both the magnetic and the capillary force scale like $\sim D^6$.
Below this limit, thermal agitation start playing a significant role and affecting the translational and rotational degrees of freedom of the particles. 
The typical particle size below which thermal agitation could free a particle from a magnetocapillary bond was estimated in~\cite{lagubeau2016} to be around \SI{3}{\micro\meter}.
In addition, the magnetic order will be greatly influenced by the thermal noise and the question of the presence of the permanent magnetic moment is hard to answer. 
How the rotational regimes would be affected by a downscaling thus remains an open question.

\section{Conclusion}

To conclude, rotating fields cause magnetocapillary assemblies to rotate thanks to three distinct mechanisms.
First, individual particles rotate, similarly to a compass needle in a magnetic induction field.
This effect can be attributed to a permanent component in the magnetization of the spheres, probably linked to their fabrication process.
Because the individual particles rotate at a different speed than the whole assembly, this motion introduces multiple length and time scales which is the ideal condition for efficient micromixing.
Secondly, dipole-dipole interactions can contribute to a reorientation.
This effect is expected to dominate at vanishing excitation frequency, where the rotation of the assembly is locked with the external field.
Furthermore, dipoles on a regular polygon have no reason to rotate as their interaction energy is independent on the orientation with regards to the field.
This means that this effect requires a deformation of the assembly.
Under high frequency and high amplitude, a very fast rotation is observed that cannot be explained by the previous two mechanisms.
In this regime characterized by high deformations, the rotation speed is proportional to the area of the cycles in the configuration space, similarly to how nonreciprocal deformations power the translational motion of microswimmers.
This fast mode is particularly suitable for propulsion, and could serve as the basis for efficient ciliate-like locomotion in more complex geometries.

\section*{Conflicts of interest}
There are no conflicts of interest to declare.

\section*{Acknowledgements}
This work was financially supported by FNRS PDR grant T.0129.18, as well as the DFG Priority Programme SPP 1726 ``Microswimmers - From Single Particle Motion to Collective Behaviour'' .
GG thanks FRIA for financial support.
Computational resources have been provided by the Consortium des \'{E}quipements de Calcul Intensif (C\'{E}CI), funded by the Fonds de la Recherche Scientifique de Belgique (F.R.S.-FNRS) under Grant No. 2.5020.11.



\balance


\bibliography{biblio} 
\bibliographystyle{rsc} 

\end{document}